\documentclass[aps,prl,floatfix,showpacs,showkeys,twocolumn,superscriptaddress]{revtex4-1}
\newcommand{\sig}{\mbox{\boldmath{$\sigma$}}}
\usepackage{graphicx}
\input{epsf}

\begin{document}

\title{Linearity of the edge states energy spectrum in the 2D topological insulator}

\author{M.V. Entin}
 \affiliation{Rzhanov Institute of Semiconductor Physics, Siberian Branch of the Russian Academy of Sciences, \\Novosibirsk, 630090, Russia}
 \affiliation{Novosibirsk State University, Novosibirsk, 630090, Russia}

\author{M.M. Mahmoodian}
 \email{mahmood@isp.nsc.ru}
 \affiliation{Rzhanov Institute of Semiconductor Physics, Siberian Branch of the Russian Academy of Sciences, \\Novosibirsk, 630090, Russia}
 \affiliation{Novosibirsk State University, Novosibirsk, 630090, Russia}

\author{L.I. Magarill}
 \affiliation{Rzhanov Institute of Semiconductor Physics, Siberian Branch of the Russian Academy of Sciences, \\Novosibirsk, 630090, Russia}
 \affiliation{Novosibirsk State University, Novosibirsk, 630090, Russia}

\date{\today}

\begin{abstract}
Linearity of the topological insulator edge state spectrum plays the crucial role for various transport phenomena. The previous studies found that this linearity exists near the spectrum crossing point, but did not determine how perfect the linearity is. The purpose of the present study is to answer this question in various edge states models. We examine Volkov and Pankratov (VP) model \cite{volk} for the Dirac Hamiltonian and the model of \cite{zhou,shen} (BHZ1) for the Bernevig, Hughes and Zhang (BHZ) Hamiltonian \cite{bern} with zero boundary conditions. It is found that both models yield ideally linear edge states. In the BHZ1 model the linearity is conserved up to the spectrum ending points corresponding to the tangency of the edge spectrum with the boundary of 2D states. In contrast, the model of \cite{vova} (BHZ2) with mixed boundary conditions for BHZ Hamiltonian and the 2D tight-binding (TB) model from \cite{bern} yield weak non-linearity.
\end{abstract}

\pacs{73.21.Fg, 73.61.Ga, 73.22.Gk, 72.25.Dc}

\maketitle

\section{Introduction}

The most important feature of 2D topological insulators (TI) is the presence of edge states overlapping the gap in 2D states. Just these states have a crossing point in the vicinity of which the spectrum is linear. This spectrum is spin-polarized. That means that the direction of motion is rigidly connected to the electron spin.

This linearity is well fulfilled in the vicinity of the conic point. The presence of linearity is reflected in a number of physical effects. For example, we can refer the reader to the electron-electron collisions in graphene \cite{we_graphene} where the linearity permits the collisions for collinear moving electrons only. In this case small non-linear corrections to the graphene spectrum play a crucial role in e-e collisions, permitting or forbidding the process near the collinear motion depending on the sign of small nonlinearity. The same arguments are true for the exciton existence in graphene \cite{we_exciton}.

The other group of phenomena affected by the linearity is thermodynamic properties. The linear edge states constant DOS results in pure linear temperature dependence of the edge states specific heat. Besides, the spectrum linearity leads to the independence of the quantum capacitance of the edge states on the electron concentration \cite{brag}.

In the 1D system of edge states the linearity yields unusual properties of transport. In fact, classical interacting electrons with such a spectrum move with the same or opposite velocities. The result is the absence of e-e scattering. In a separate paper we shall show that the quantum many-body problem is exactly solvable and gives just the same picture as a classical one.

Generally speaking, the nonlinear corrections to the linear crossing spectrum can be obtained by the k-p expansion on the quasimomentum around the crossing (conic) point. However, the symmetry relations can forbid the corresponding contributions. Hence, it is important to clarify, how strong is the linearity of the TI edge electron spectrum? This is the purpose of the present paper. The system under examination is the most popular one: namely, a HgTe layer in the CdTe matrix (see, e.g. \cite{konig2,kane,kvon1,kvon2,Roth,shen}). In cases when the numerical calculations are needed, we use the numerical parameters of this system.

First, we shall consider the VP model \cite{volk}. In this model, quite straightforwardly, it is followed that the electron edge states have an exact linear spectrum. This conclusion does not depend on the abruptness of the transition domain. We, however, carefully examine this model as applied to 2D TI (original version is 3D TI \cite{volk}). The problem with this model is e-h symmetry and only 2 bands are included in the consideration. The question arises, is the linearity connected with the model assumptions, or it has more general character (for example, conditioned by topological reasons)? It seems that the k-p method applied to the 1D Hamiltonian of edge states would yield the corrections to the spectrum of higher orders in the 1D momentum of the edge state. Hence, then we will study the spectrum linearity in other different known models: BHZ1 \cite{zhou,shen}, BHZ2 \cite{vova} and TB \cite{bern}.

Further, we shall study the BHZ1 model, the best known model of 2D TI edge states. This model is based on the 4x4 Hamiltonian, non-symmetric with respect to the e-h replacement. We also conclude that the linearity of two edge states is conserved in this model. Besides, we will deal with the BHZ2 model \cite{vova} which is the generalization of BHZ1 model. In the BHZ2 model the mixed boundary condition is used instead of the zero boundary condition in the BHZ1 model. Here we shall prove that the linearity is approximate, but numerically good enough.  Then we examine the tight-binding square lattice model (TB) \cite{bern} with 4 states in an elementary cell. It was found that, in this case, the spectrum also has small non-linear corrections.

\section{k-p arguments}

The states near the conic point can be considered by means of the k-p perturbation theory. Let us have a 2D crystal $y>0$ with the edge $y=0$. Due to the periodicity in the $x$ direction, the Bloch states can be ascribed by continuous $x$-momentum $p_x$ and discrete (for edge states) or other continuous (for 2D states) quantum number. In the spirit of the k-p theory, we should take the basis of the states at $p_x=0$ and formulate the Hamiltonian at a finite $p_x$ as a matrix in the basic states. Let there be a single double-degenerate edge state at $p_x=0$ and continuous states numerated by the $p_y>0$ momentum. The Hamiltonian can be written as $H=\epsilon_0+p_x\hat{P}_x/m_0+p_x^2/2m_0$, where $m_0$ is the free electron mass. Operator $\hat{P}_x$ is determined by the matrix over the states at zero momentum $p_x$. The linear approximation is given by the block of this matrix between the degenerate edge states $\sigma$ $p_x(\hat{P}_x)_{\sigma,\sigma'}/m_0$ (if, indeed, these matrix elements are non-vanishing). This gives the linear spectrum near point $p_x=0$. However, the second-order approximation in $p_x$ yields the second-order corrections to the Hamiltonian matrix in degenerate states:
\begin{equation}\label{h1}
H^{(2)}_{\sigma,\sigma'}=\frac{p_x^2}{2m_0}\delta_{\sigma,\sigma'}+\frac{p_x^2}{m_0^2}\sum_{p_y,\nu}
\frac{(\hat{P}_x)_{\sigma;p_y,\nu}(\hat{P}_x)_{p_y,\nu;\sigma'}}{\epsilon_0-\epsilon(p_x=0,p_y;\nu)}.
\end{equation}
Here $\nu$ are all discrete numbers, characterizing the 2D states, $\epsilon(p_x,p_y;\nu)$ is the 2D spectrum. Eq.~(\ref{h1}) shows that, in general, the quadratic corrections to the spectrum do not vanish. The exception may be caused by some group prohibitions. However, below, we shall see that this is just the case in the the VP and BHZ1 models.

\section{VP model}

In the 2D case the Hamiltonian of the VP model has the form
\begin{equation}\label{hh}
   H=\left(
       \begin{array}{cc}
         \Delta\sigma_0 &v\sig{\bf p} \\
         v\sig{\bf p} & -\Delta\sigma_0 \\
       \end{array}
     \right),
\end{equation}
where ${\bf p}=(p_x,p_y) $ is the 2D momentum operator, $\sigma_0$ is the $2\times2$ unit matrix, $\sig=(\sigma_x,\sigma_y,\sigma_z)$ are the Pauli matrices.

Let $\Delta=\Delta(y)$ and $\Delta(y)>0$ at $y<0$, $\Delta(y)<0$ at $y>0$. The Schr\"odinger equation $H\chi=\epsilon\chi$ has the solutions localized in $y$-direction
\begin{equation}\label{chi}
    \chi_\sigma=\frac{\chi_\sigma^{(0)}e^{ip_xx}}{\sqrt{8{\cal N}L_x}}\exp\left(-\frac{1}{v}\int\limits_0^y\Delta(y')dy'\right),
\end{equation}
\begin{equation}\label{norm}
\chi_\sigma^{(0)}=\left(
           \begin{array}{c}
            1-\sigma \\
            1+\sigma \\
            1+\sigma \\
            \sigma-1 \\
           \end{array}
         \right),~~~
{\cal N}=\int\limits_{-\infty}^{\infty}dy\exp\left(-\frac{2}{v}\int\limits_0^y\Delta(y')dy'\right),
\end{equation}
with energies $\epsilon_\sigma=\sigma vp_x$, where $\sigma=\pm1$ is a spin quantum number, $L_x$ is the edge length. The edge states decay from the line $y=0$. The linear spectrum has no corrections of higher order in momentum $p_x$.

In a particular case of a step-like behaviour of $\Delta(y)=\Delta_+\theta(y)+\Delta_-\theta(-y)$, where $\Delta_+<0$, $\Delta_->0$,
\begin{equation}\label{chi2}
    \chi_\sigma=\frac12\sqrt{\frac{\Delta_+\Delta_-}{v(\Delta_+-\Delta_-)L_x}}\chi_\sigma^{(0)}e^{ip_xx}\exp\left(\frac{\Delta(y)y}{v}\right).
\end{equation}

\begin{figure}[ht]
\leavevmode\centering{\epsfxsize=6cm\epsfbox{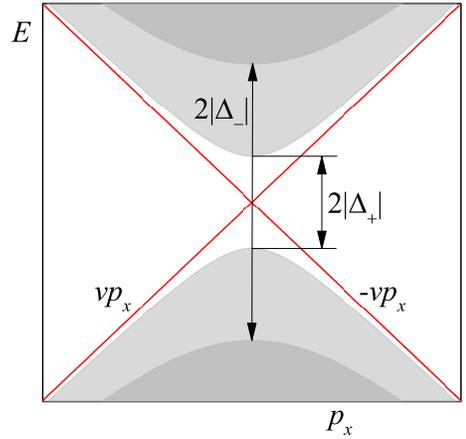}}
\caption{(Color online). Edge spectrum of TI in the step-like VP model for $|\Delta_+|<|\Delta_-|$. The domains of 2D states of TI and ordinary insulator are filled. At $p_x\to\pm \infty$, edge branches $\pm vp_x$ asymptotically approach the 2D states domain boundaries $\sqrt{\Delta_\pm^2+p_x^2}$ and $-\sqrt{\Delta_\pm^2+p_x^2}$.}\label{fig1}
\end{figure}

Thus, we come to the statement that the linearity of the surface states in this model is {\it absolute}. The linear edge branches overlap in energy with the 2D TI spectrum and have no ending points (see Fig.~\ref{fig1}). The situation with BHZ Hamiltonian is different.

\section{BHZ Hamiltonian with zero boundary conditions}

In this section we present an exact solution of the edge states for BHZ Hamiltonian \cite{bern} with the zero boundary condition \cite{shen}. The effective 4x4 BHZ Hamiltonian was derived  in \cite{bern} from the Kane model for a quantum well of HgTe/CdTe:
\begin{equation}\label{bhz}
   H_{BHZ}=\left(
       \begin{array}{cc}
         h(\bf{p}) & 0 \\
         0 & h^*(-\bf{p}) \\
       \end{array}
     \right),
\end{equation}
where $h(\bf{p})=\epsilon({\bf p})\sigma_0+{\bf d}({\bf p})\sig$, $\epsilon({\bf p})=-Dp^2$, ${\bf d}({\bf p})=(Ap_x,Ap_y,M-Bp^2)$. $A, B, D$ and $M$ are the parameters determined by the quantum well thickness and the material parameters.

For the edge states, $p_y$ is imaginary, $p_y=-i\lambda$ and the 2D solutions of the Schr\"{o}dinger equation can be searched in the form of $\chi\propto e^{ip_x-\lambda y}$ ($\lambda>0$). The Hamiltonian (\ref{bhz}) is a block diagonal, and the eigenvalue problem of the upper and lower blocks can be solved separately. Thus, we should solve Schr\"{o}dinger equations $h({\bf p})\psi_{+1}=E\psi_{+1}$ or $h^*(-{\bf p})\psi_{-1}=E\psi_{-1}$, where $\psi_\sigma$ are two-component spinors. The characteristic equations  $\det[h({\bf p})-E]=0$,  $\det[h^*(-{\bf p})-E]=0$ have the same real roots $\pm\lambda_1$ and $\pm\lambda_2$,
\begin{eqnarray}\label{roots}
\lambda_{1,2}^2=p_x^2+F\pm\sqrt{F^2-\frac{M^2-E^2}{B_+B_-}},
\end{eqnarray}
where $B_\pm=B\pm D$, $F=[A^2-2(ED+BM)]/(2B_+B_-)$, $\lambda_{1,2}>0$. To satisfy the boundary condition $\psi_\sigma|_{y=0}=0$ we combine the solutions with $\lambda_1$ and $\lambda_2$
\begin{eqnarray}\label{wf}
\psi_\sigma=
\left(\begin{array}{c}
                         a \\
                         b
                       \end{array}
               \right)
\left(e^{-\lambda_1 y}-e^{-\lambda_2 y}\right).
\end{eqnarray}
The substitution to the Schr\"odinger equation yields:
\begin{eqnarray}\label{disp1}
ED+BM-B_+B_-\left(p_x^2+\lambda_1\lambda_2\right)=0,
\end{eqnarray}
which is solvable if:
\begin{eqnarray}\label{cond1}
ED+BM-B_+B_-p_x^2>0.
\end{eqnarray}
Eq.~(\ref{disp1}) has exact solutions
\begin{eqnarray}\label{en1}
E_\sigma=-M\frac{D}{B}+\sigma A\sqrt{\frac{B_+B_-}{B^2}} p_x.
\end{eqnarray}
The condition (\ref{cond1}) for the Eq.~(\ref{en1}) gives $p_{x,\sigma}^{(l)}<p_x<p_{x,\sigma}^{(r)}$, where
\begin{eqnarray}\label{pxmax}
p_{x,\sigma}^{(l,r)}=\frac{-\sigma AD\pm\sqrt{A^2D^2+4BMB_+B_-}}{2B\sqrt{B_+B_-}}.
\end{eqnarray}
Limits $p_{x,\sigma}^{(l,r)}$ play the role of the linear spectrum ending points. In these points, the linear dependence touches the boundary of 2D states for valence or conduction bands (corresponding to $p_y=0$). Note, that in contrast to~\cite{shen}, solutions (\ref{en1}) with the linear dependence on $p_x$ are {\it exact}. Both exact linearity and the presence of ending points have been unknown up to now.

\begin{figure}[ht]
\leavevmode\centering{\epsfxsize=7.5cm\epsfbox{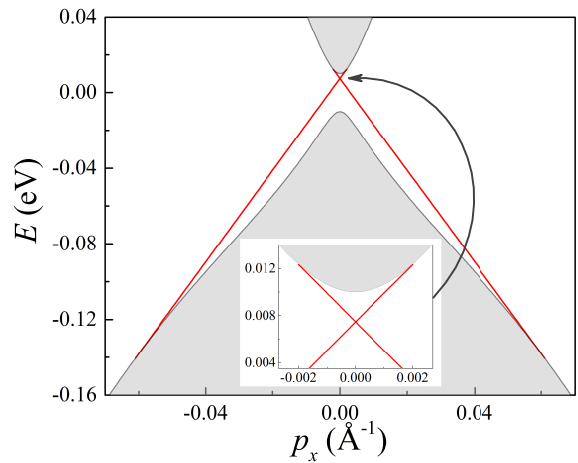}}
\caption{(Color online). The edge states spectrum in the BHZ1 model. The crossed tilted straight lines correspond to linear branches. The domain of 2D states is filled. The linear branches terminate in the tangency points to the 2D spectrum border.}\label{fig2}
\end{figure}

We have made calculations with the use of the CdTe/\-HgTe/CdTe system parameters from \cite{qi}. We consider the 7 nm width quantum well for which $A=3.645$ eV$\cdot\mbox{\AA}$, $B=-68.6$ eV$\cdot\mbox{\AA}^2$, $D=-51.2$ eV$\cdot\mbox{\AA}^2$, $M=-0.010$~eV. (The same parameters will be used in the numerical calculations below). The results, obtained by means of Eqs. (\ref{en1}) and (\ref{cond1}), are presented in Fig.~\ref{fig2}. One can see that the spectrum remains linear up to the ending points, where the branches are tangent to the border of the 2D spectrum.

\section{BHZ Hamiltonian with a mixed boundary condition}

In this section we use the same Hamiltonian (\ref{bhz}), but with the mixed boundary conditions accounting for the admixture of higher crystal bands via boundary condition parameters. The general form of mixed boundary conditions was derived in \cite{vova}. We shall consider the case when boundary potential does not mix the spin of electrons,
\begin{eqnarray}\label{bcgen}
\left(
    \begin{array}{cc}
        \frac{1}{q}\partial_y + 1 & 0 \\
        0 & 1 \\
    \end{array}
\right)\psi_\sigma|_{y=0}=0.
\end{eqnarray} $q=gA/B_+$. At $g\to-\infty$ we return to the case of zero boundary. The wave function satisfying the boundary condition (\ref{bcgen}) has the form:
\begin{equation}\label{wfgen}
\psi_\sigma=
\left(
    \begin{array}{c}
        a\left(\frac{e^{-\lambda_1 y}}{q-\lambda_1}-\frac{e^{-\lambda_2 y}}{q-\lambda_2}\right) \\
        b\left(e^{-\lambda_1 y}-e^{-\lambda_2 y}\right)
    \end{array}
\right).
\end{equation}
The substitution to the Schr\"odinger equation yields:
\begin{eqnarray}\label{dispgen1}
ED+BM-B_+B_-\left(p_x^2+\lambda_1\lambda_2\right)-A^2\gamma/2=0,\\
\gamma=\frac{2B_+(\sigma q-p_x)p_x+M-E}{B_+(q^2-p_x^2)+M-E}.
\end{eqnarray}
The domain of permitted values of the longitudinal momentum $p_x$ is given by an inequality
\begin{eqnarray}\label{condgen1}
ED+BM-B_+B_-p_x^2-A^2\gamma/2>0.
\end{eqnarray}

\begin{figure}
\leavevmode\centering{\epsfxsize=7.5cm\epsfbox{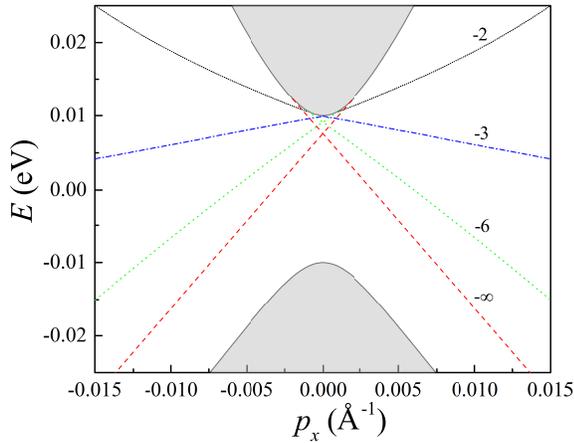}}
\caption{(Color online). BHZ2 model: Evolution of the edge states spectrum with boundary condition parameter $g$ (marked at the curves). The upper and lower curves restrict the domain of 2D states. The curves touch the 2D spectrum where they terminate.}\label{fig3}
\end{figure}

Eq.~(\ref{dispgen1}) can be solved approximately when $g$ tends to $-\infty$. In the main order on $1/g$ we obtain
\begin{eqnarray}\label{engen}
E_\sigma\approx-M\frac{D}{B}+\sigma A\sqrt{\frac{B_+B_-}{B^2}} p_x\nonumber\\
+\frac{B_+\sqrt{B_+B_-}}{gB^2}\left(-M+\frac{\sigma ADp_x}{\sqrt{B_+B_-}}+Bp_x^2\right).
\end{eqnarray}
The evolution of the edge states spectrum with boundary condition parameter $g$ is demonstrated in Fig.~\ref{fig3}. The case of $g=-\infty$ corresponds to the zero boundary condition. In contrast to the BHZ1 model, the edge states get a small, but noticeable curvature.

\section{Tight-binding model}

The square lattice with four states per unit cell, introduced in \cite{bern}, is another representative of topological insulators. We examine this system to check the linearity of the edge spectrum.

The Hamiltonian of infinite lattice can be obtained from Eq.~(\ref{bhz}) by the replacements: $\epsilon({\bf p})=-2Da^{-2}[2-\cos(p_xa)-\cos(p_ya)]$, ${\bf d}({\bf p})=(Aa^{-1}\sin(p_xa),Aa^{-1}\sin(p_ya),M-2Ba^{-2}[2-\cos(p_xa)-\cos(p_ya)])$. Here $a$ is the lattice constant. In half-infinite lattice limited in $y$-direction the lattice Hamiltonian can be reduced to infinite problem with zero boundary condition for the wave function on the atomic row preceding to the first. Hence, we will base on the envelope function approach with the boundary condition $\psi_\sigma|_{y=0}=0$, where $y=0$ is chosen for this row.

The characteristic equation have real roots $\pm\lambda_1$ and $\pm\lambda_2$,
\begin{eqnarray}\label{roots-tight}
\cosh(\lambda_{1,2}a)=2-\cos(p_xa)+G\pm\sqrt{G^2-K},\nonumber\\
G=\frac{A^2\left[2-\cos(p_xa)\right]-2(ED+BM)}{4B_+B_-a^{-2}-A^2},\\
K=\frac{(M^2-E^2)a^2-2A^2\left[1-\cos(p_xa)\right]^2}{4B_+B_-a^{-2}-A^2}.\nonumber
\end{eqnarray}
To satisfy the boundary condition $\psi_\sigma|_{y=0}=0$, we will use the same wave function as in Eq.~(\ref{wf}). The substitution to the Schr\"odinger equation yields:
\begin{eqnarray}\label{disp-tight}
ED+BM-\frac{2B_+B_-}{a^2}\bigg(2-\cos(p_xa)~~~~~~~~~~~~~~~~~\\
-\frac{\sinh([\lambda_1-\lambda_2]a)}{\sinh(\lambda_1a)-\sinh(\lambda_2a)}\bigg)=0.\nonumber
\end{eqnarray}
The domain of permitted values of the longitudinal momentum $p_x$ is given by an inequality
\begin{eqnarray}\label{cond-tight}
ED+BM-2B_+B_-a^{-2}\big(1-\cos(p_xa)\big)>0.
\end{eqnarray}

The energy spectrum was found numerically according to Eq.~(\ref{disp-tight}) with different values of lattice constant $a$. For a small $a$ the edge spectrum corresponds to the case of BHZ1. For a large $a$ the edge branches obtain almost unnoticeable curvature. The ending points of these branches coincide with the tangential points to the 2D spectrum border, as in BHZ1 and BHZ2 models.

\begin{figure}
\leavevmode\centering{\epsfxsize=7.5cm\epsfbox{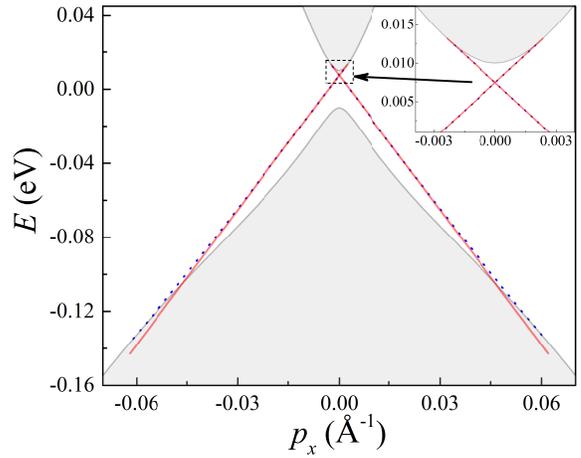}}
\caption{\label{fig4} (Color online). The edge states in the 2D tight-binding model for the lattice constant $a=8\mbox{\AA}$ (dots) in comparison with the linear dispersion (lines). Inserts show magnified domains near crossing point and near the spectrum ending point.}
\end{figure}

\subsection*{Discussion and conclusions.}

As we see from the foregoing, the results of different models differ. It is important to determine which one is consistent with real or possible situations. We have both pro and contra arguments of each of these models.

The VP model is based on the simplest 2D k-p Hamiltonian. To be valid this model implies the absence of admixture of other bands except for the two considered and the smooth behavior of function $\Delta(y)$. In fact, such situation can be the case if the edge of the TI is overetched or enriched by some ions during the etching process. For example, the TI can be thinned towards its edges: that in the case of HgTe, will lead to a change of the gap sign. Such process would lead to smooth $\Delta(y)$. At the same time, the quadratic terms in the Hamiltonian, caused by the far bands of the 3D crystal (distant by some eV), can be very small, as compared with scale of 20 meV bandgap of the TI. The other property of the VP model is the presence of the wave functions tail outside the TI domain. This tail becomes short if the gap in the insulator domain grows. The wave function does not vanish at the boundary. Instead, it has a maximum (nib) there. In the limit of a large gap in the insulating part, such wave function satisfies the boundary condition consistent with (and determined by) the Schr\"odinger equation for the internal part of TI.

The BHZ1 model has the advantage that it is connected with the real 2D structure of the 2D quantum well. At the same time, the edge of this system is described roughly: the real structure at the edge and the far bands are not taken into account. The last circumstance is accounted for in \cite{vova}, but the boundary condition of the general form in \cite{vova} does not take into account the limitation caused by the Schr\"odinger equation in the internal part of TI. This becomes apparent in comparison with the VP model.

And finally, the tight-binding 2D model describes the HgTe quantum well only nominally. However, it can be applied to the artificial lattice with such structure.

Thus, we conclude that all the considered models have both advantages and disadvantages.

In conclusions, we have proved the exact linearity of the edge states in the VP and BHZ1 models and the approximate linearity in the BHZ2 and tight-binding models of 2D TI. The edge states overlap with the 2D states. The linearity propagates up to the ending points of the edge state spectrum, where the linear spectrum is tangent to the border of the 2D spectrum. Thus, we conclude that the linearity of the edge states energy spectrum is conserved in the most part of the interesting momentum domain.

In a separate paper, with the authorship of one of the authors (MVE), the consequences of spectrum linearity will be examined for a system of interacting electrons. In advance, it can be announced that these consequences drastically change the interaction character, as compared to the quadratic spectrum.

It should be emphasized that the overlapping of the edge states with the 2D states continuum makes the elastic exchange of carriers between these states possible. This process, which we hope to study later, should play a determinative role in the electron kinetics. If this exchange exists (even very slow), the transport can not be studied separately as 2D and edge components, but should be considered solving the related equations.

\section*{Acknowledgements}

This research was supported by RFBR grant No 17-02-00837.


\begin{thebibliography}{20}

\bibitem{volk} B.A. Volkov and O.A. Pankratov, JETP Lett. {\bf 42}, 178 (1985).
\bibitem{zhou} Bin Zhou, Hai-Zhou Lu, Rui-Lin Chu, Shun-Qing Shen, and Qian Niu, Phys.Rev.Lett. {\bf 101}, 246807 (2008).
\bibitem{shen} S.-Q. Shen, Topological insulators, Springer Series in Solid-State Sciences {\bf 174} (Springer, New York) (2012).
\bibitem{bern} B.Andrei Bernevig, Taylor L.Hughes, Shou-Cheng Zhang, Science {\bf 314}, no. 5806, 1757 (2006).
\bibitem{vova} V.V. Enaldiev, I.V. Zagorodnev, V.A Volkov, JETP Lett. {\bf 101}, 89 (2015).
\bibitem{we_graphene} L.E. Golub, S.A. Tarasenko, M.V. Entin, and L.I. Magarill, Phys. Rev. B {\bf 84}, 195408 (2011).
\bibitem{we_exciton} M.M. Mahmoodian and M.V. Entin, EPL, {\bf 102}, 37012 (2013).
\bibitem{brag} L.S. Braginsky and M.V. Entin, arXiv:1703.04231 [cond-mat].
\bibitem{konig2} M. K\"{o}nig, S. Wiedmann, C. Brune, A. Roth, H. Buhmann, L.W. Molenkamp, X.-L. Qi, and S.-C. Zhang, Science, {\bf 318}, 766 (2007).
\bibitem{kane} C.L. Kane and E.J. Mele, Phys. Rev. Lett., {\bf 95}, 146802 (2005).
\bibitem{kvon1} G.M. Gusev, Z.D. Kvon, O.A. Shegai, N.N. Mikhailov, S.A. Dvoretsky, and J.C. Portal, Phys. Rev. B {\bf 84}, 121302 (R) (2011).
\bibitem{kvon2} G.M. Gusev, Z.D. Kvon, E.B. Olshanetsky, A.D. Levin, Y. Krupko, J.C. Portal, N.N. Mikhailov, and S.A. Dvoretsky, Phys. Rev. B. {\bf 89}, 125305, (2014).
\bibitem{Roth} A. Roth, C. Brune, H. Buhmann, L.W. Molenkamp, J. Maciejko, Xiao-Liang Qi, Shou-Cheng Zhang, Science, {\bf 325}, no. 5938, 294 (2009).
\bibitem{qi} Xiao-Liang Qi, Shou-Cheng Zhang, Rev. Mod. Phys. {\bf 83}, 1057 (2011).

\end{thebibliography}
\end{document}